\documentclass[10pt,A4paper,twocolumn,aps,pra, superscriptaddress,longbibliography, nofootinbib]{revtex4-2}

\usepackage{amsmath}    
\usepackage{graphicx}   
\usepackage{color}      
\usepackage[usenames,dvipsnames,svgnames,table]{xcolor}
\usepackage[hidelinks]{hyperref}   
\usepackage{float}
\usepackage{braket}
\usepackage{bbm}
\usepackage{multibib}
\newcites{meth}{Methods References}

\usepackage[capitalise]{cleveref}
\Crefname{figure}{Fig.}{Figs.}
\Crefname{table}{Tab.}{Tabs.}
\usepackage{siunitx}
\usepackage{amssymb}

\usepackage{rotating}
\usepackage{tabularx}
\usepackage{cellspace}
\usepackage{booktabs}
\usepackage{graphics}
\usepackage{newfloat}
\usepackage{soul}
\newcolumntype{Y}{>{\centering\arraybackslash}X}
\newcolumntype{d}{c}
\newcolumntype{a}{c}

\usepackage{xr}

\usepackage{savesym}
\savesymbol{printendnotes}
\usepackage{enotez}
\restoresymbol{NT}{printendnotes}
\newcommand{\emptymark}[1]{}
\setenotez{list-name=Methods, mark-cs=\emptymark}

\usepackage{xpatch}
\makeatletter
\xpatchcmd{\@ssect@ltx}{\@xsect}{\protected@edef\@currentlabelname{#8}\@xsect}{}{}
\xpatchcmd{\@sect@ltx}{\@xsect}{\protected@edef\@currentlabelname{#8}\@xsect}{}{}
\makeatother

\definecolor{adpcolor}{rgb}{0.36,0.54,0.66}

\begin{document}

\title{Calibration of Drive Non-Linearity for Arbitrary-Angle \\ Single-Qubit Gates Using Error Amplification}

\author{Stefania~Laz\u{a}r}
\thanks{These authors contributed equally to this work.}
\affiliation{Department of Physics, ETH Zurich, CH-8093 Zurich, Switzerland}

\author{Quentin~Ficheux}
\thanks{These authors contributed equally to this work.}
\affiliation{Department of Physics, ETH Zurich, CH-8093 Zurich, Switzerland}

\author{Johannes~Herrmann}
\affiliation{Department of Physics, ETH Zurich, CH-8093 Zurich, Switzerland}

\author{Ants~Remm}
\affiliation{Department of Physics, ETH Zurich, CH-8093 Zurich, Switzerland}

\author{Nathan~Lacroix}
\affiliation{Department of Physics, ETH Zurich, CH-8093 Zurich, Switzerland}

\author{Christoph~Hellings}
\affiliation{Department of Physics, ETH Zurich, CH-8093 Zurich, Switzerland}

\author{Francois~Swiadek}
\affiliation{Department of Physics, ETH Zurich, CH-8093 Zurich, Switzerland}

\author{Dante~Colao~Zanuz}
\affiliation{Department of Physics, ETH Zurich, CH-8093 Zurich, Switzerland}

\author{Graham~J.~Norris}
\affiliation{Department of Physics, ETH Zurich, CH-8093 Zurich, Switzerland}

\author{Mohsen~Bahrami~Panah}
\affiliation{Department of Physics, ETH Zurich, CH-8093 Zurich, Switzerland}
\affiliation{ETH Zurich - PSI Quantum Computing Hub, Paul Scherrer Institute, CH-5232 Villigen, Switzerland}

\author{Alexander~Flasby}
\affiliation{Department of Physics, ETH Zurich, CH-8093 Zurich, Switzerland}
\affiliation{ETH Zurich - PSI Quantum Computing Hub, Paul Scherrer Institute, CH-5232 Villigen, Switzerland}

\author{Michael~Kerschbaum}
\affiliation{Department of Physics, ETH Zurich, CH-8093 Zurich, Switzerland}
\affiliation{ETH Zurich - PSI Quantum Computing Hub, Paul Scherrer Institute, CH-5232 Villigen, Switzerland}

\author{Jean-Claude~Besse}
\affiliation{Department of Physics, ETH Zurich, CH-8093 Zurich, Switzerland}
\affiliation{ETH Zurich - PSI Quantum Computing Hub, Paul Scherrer Institute, CH-5232 Villigen, Switzerland}

\author{Christopher~Eichler}
\affiliation{Department of Physics, ETH Zurich, CH-8093 Zurich, Switzerland}

\author{Andreas~Wallraff}
\affiliation{Department of Physics, ETH Zurich, CH-8093 Zurich, Switzerland}
\affiliation{ETH Zurich - PSI Quantum Computing Hub, Paul Scherrer Institute, CH-5232 Villigen, Switzerland}
\affiliation{Quantum Center, ETH Zurich, 8093 Zurich, Switzerland}

\date{\today}

\begin{abstract}
    The ability to execute high-fidelity operations is crucial to scaling up quantum devices to large numbers of qubits. However, signal distortions originating from non-linear components in the control lines can limit the performance of single-qubit gates. In this work, we use a measurement based on error amplification to characterize and correct the small single-qubit rotation errors originating from the non-linear scaling of the qubit drive rate with the amplitude of the programmed pulse. With our hardware, and for a 15-ns pulse, the rotation angles deviate by up to several degrees from a linear model. Using purity benchmarking, we find that control errors reach $2\times 10^{-4}$, which accounts for half of the total gate error. Using cross-entropy benchmarking, we demonstrate arbitrary-angle single-qubit gates with coherence-limited errors of $2\times 10^{-4}$ and leakage below $6\times 10^{-5}$. While the exact magnitude of these errors is specific to our setup, the presented method is applicable to any source of non-linearity. Our work shows that the non-linearity of qubit drive line components imposes a limit on the fidelity of single-qubit gates, independent of improvements in coherence times, circuit design, or leakage mitigation when not corrected for.
\end{abstract}

\maketitle

\section{INTRODUCTION}

As quantum processors accommodate an ever larger number of qubits, understanding the errors that limit their performance remains an important task to enable further progress~\cite{Blais2021,Kjaergaard2020a,Wendin2017}. In particular, errors in single- and two-qubit gates limit the fidelities of quantum error correction codes~\cite{Kelly2015,Andersen2020b,Marques2021c,Krinner2022, Chen2021p}, variational quantum algorithms~\cite{Kandala2017,Lacroix2020}, or digital quantum simulations~\cite{Salathe2015,Barends2016,Arute2020b}. These errors can be reduced by increasing the coherence times, but also by mitigating control errors~\cite{McKay2019,Rudinger2019,Sarovar2020,McKay2020,Krinner2020,Barends2019} or by reducing the circuit depth when compiling algorithms with a larger gate set~\cite{Foxen2020a, Lacroix2020, Herrmann2022}.

For superconducting quantum processors, standard gate calibration techniques~\cite{Chen2016,Kelly2018, Krantz2019,Xu2021b} assume that the control parameters respond linearly to the control fields. However, commonly used electronic components can introduce non-linear effects, which result in control errors if unaccounted for. For example, IQ-mixer imperfections produce spurious frequency components~\cite{Herrmann2022a,Jolin2020,Xu2021j}, gain compression in amplifiers saturates the power of control signals~\cite{Pedro2018}, and saturation of pulse-generation devices lead to waveform distortions~\cite{Chaves2021}.

For the past decade, two-qubit gates have imposed the main limitation on superconducting quantum processors~\cite{Kjaergaard2020a,Gambetta2017}. However, considerable efforts in recent years have reduced the errors of two-qubit gates into the range of $10^{-3}$~\cite{Negirneac2021,Ficheux2021,Barends2019,Sung2021a,Kandala2021,Wei2021a}, making their contribution to computation errors comparable to that of single-qubit gates~\cite{Blais2021}. Moreover, current algorithms often require a larger number of single-qubit gates than two-qubit gates, especially for dynamical decoupling schemes~\cite{Bylander2011,Carr1954}. Thus, further improvements in the performance of single-qubit gates are needed to enhance the overall fidelity of quantum algorithms.

Arbitrary single-qubit rotation angles are routinely calibrated by assuming that the rotation of the state vector scales linearly with the pulse amplitude~\cite{Chen2016,Kelly2018,Krantz2019,Xu2021b}. However, non-linear electronic components in the drive line can produce a deviation from this linear model that leads to rotation errors. So far, few techniques have been developed to compensate for similar imperfections, such as waveform predistortion~\cite{Gustavsson2013,Rol2020,Negirneac2021}, pulse-shape optimization methods~\cite{Werninghaus2021,Singh2022c}, and correction of drive-pulse amplitudes affected by imperfect signal generation~\cite{Chaves2021}. 

Here, we present an \emph{in-situ} amplification scheme for small rotation errors, which allows to calibrate coherence-limited arbitrary-angle single-qubit gates on a transmon qubit~\cite{Koch2007} even in the presence of non-linearity. With this technique, we correct for residual rotation errors when using pulses with amplitudes approaching the 1-dB compression point of the frequency conversion device in our microwave line, which is non-linear at high input powers~\cite{Pedro2018}. Using randomized benchmarking (RB)~\cite{Magesan2011}, purity benchmarking (PB)~\cite{Wallman2015a}, and cross-entropy benchmarking (XEB)~\cite{Boixo2018a}, we demonstrate arbitrary-angle 15-ns-long single-qubit gates with coherence-limited performance and with leakage outside the computational subspace below $6\times 10^{-5}$. Importantly, we find that, for short gate duration, coherent errors in the computational subspace caused by non-linear pulse amplitude distortions are one order of magnitude larger than leakage, imposing a lower bound on the gate error for a given gate duration, independent of improvements in coherence times.

We study the effect described above on two flux-tunable superconducting transmon qubits from two different 17-qubit devices labeled A and B, with similar designs to the one discussed in Ref.~\cite{Krinner2022}. The devices are controlled by nominally the same electronics. The measurements presented in Sec.~\ref{sec:np_calib} and \ref{sec:cont_xeb} were done on device A, while those discussed in Sec.~\ref{sec:prb} and \ref{sec:XEB} were performed on device B (see Appendix~\ref{app:qbs_used}, where we also summarize the device parameters).

\section{Pulse amplitude calibration using error amplification} \label{sec:np_calib}

\begin{figure*}[t]
  \centering
  \includegraphics{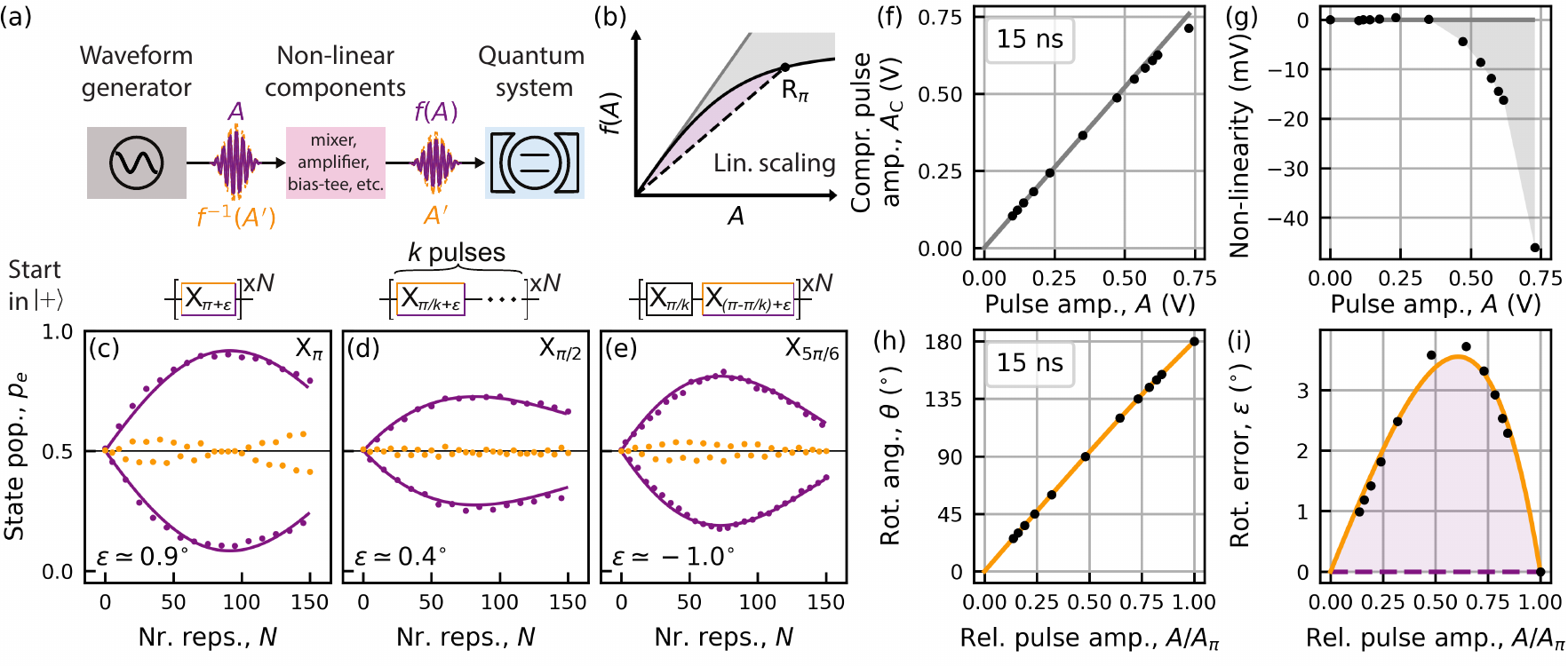}
  \caption{
  (a)~A pulse with amplitude $A$ (purple) produced by an ideal waveform generator is distorted by a series of non-linear components before reaching the quantum system with an amplitude $f(A)$, which is lower than expected. A pulse with amplitude scaled by the inverse response function $f^{-1}(A')$ ensures that the quantum system is driven with the desired amplitude $A'$.
  (b)~Schematic example of a non-linear response function $f(A)$ (solid black line) and the linear dependence (dashed line) assumed when single-qubit rotations are implemented with a linear scaling of the pulse amplitude relative to the amplitude of a $\pi$ rotation.
  (c),~(d),~(e)~Measured excited-state population of the qubit ($p_e$) after applying a pulse sequence with $N$ repetitions of the 32-ns gates shown above each panel. At the bottom left, we indicate the rotation errors $\varepsilon$ extracted from a fit of the data (purple line and points) to a master-equation-based model. The orange points are obtained after correcting the pulse amplitudes based on $\varepsilon$.
  We increment $N$ from 0 to 150 in steps of 5 in (c) and (d), and in steps of 3 in (e). 
  (f)~Compressed pulse amplitude $A_{\mathrm{C}}$ as a function of the applied pulse amplitude for a 15-ns gate. The gray line is the expected linear response. (g)~Difference between the measured data and the gray line in (f). (h)~Same data as in (f) but converted to the calibrated rotation angle $\theta$ as a function of the pulse amplitude relative to the amplitude of a $\pi$ rotation. The orange line is a fifth-order odd polynomial fit to the data. (i)~Difference of the measured data/polynomial fit in (h) and a linear interpolation between the last data point in (h) and the origin (purple line).
  }
  \label{fig:fig1}
\end{figure*}

Non-linear components in qubit control lines distort coherent microwave drive-pulses, effectively changing the amplitude of the driving field at the qubit frequency in an unintended way~\cite{Herrmann2022a}. We illustrate this effect in Fig.~\ref{fig:fig1}(a, b), where a pulse produced by the waveform generator with amplitude $A$ (purple) reaches the qubit with an amplitude which shows a reduction from a linear behavior (black and gray lines, respectively). This reduction becomes more pronounced with larger input amplitudes and leads to a rotation angle of the qubit state which is smaller than the targeted one. The pulse distortions introduced by the non-linearity could be corrected with a sample-by-sample non-linear transformation. However, we show that a simple scaling of the pulse amplitude by the inverse of the response function of the drive line (orange pulses) is enough to mitigate the rotation errors produced by the non-linearity.
 
In order to characterize the impact of non-linearity on the pulse amplitudes, we measure the rotation errors which are introduced by reducing the pulse amplitude proportionally to the amplitude of a $\pi$ rotation, see the dashed black line in Fig.~\ref{fig:fig1}(b). We use an error amplification sequence -- which we refer to as an \emph{$N$-pulse calibration} method -- inspired by the ones described in Refs.~\cite{Asaad2016,Sheldon2016a}. 

To calibrate any rotation errors left over from an inital Rabi calibration of a $\pi$ pulse, we start by initializing the qubit in the superposition state $|+\rangle = (|g \rangle + | e \rangle)/\sqrt{2}$, and then we apply $N$ repetitions of $X_{\pi+\varepsilon}$ gates, which have a small error in the rotation angle, $\varepsilon$. The pulse sequence is shown above Fig.~\ref{fig:fig1}(c). In the ideal case ($\varepsilon=0$), the qubit remains in the equatorial plane of the Bloch sphere, where the ground and excited state populations are $p_g=p_e=0.5$. For imperfect gates ($\varepsilon \neq 0$), the rotation errors accumulate to produce a deviation of the excited state population from 0.5 [see purple points in Fig.~\ref{fig:fig1}(c)]. We ignore any small errors in the initialization pulse as they will not be amplified during the pulse sequence.

We extend the methods in Refs.~\cite{Asaad2016,Sheldon2016a} by extracting the rotation error per gate from a fit of the excited-state population of the qubit to a master-equation-based model (see Appendix~\ref{app:fit}). We find around $0.9^{\circ}$ of rotation error. This corresponds to around $1.7$ mV of amplitude error for a pulse amplitude of $335 \mathrm{~mV}$ (at the output of the arbitrary waveform generator), which was obtained from an initial Rabi calibration of a 32-ns DRAG pulse~\cite{Motzoi2009,Gambetta2011a}. For the DRAG pulse, we use a Gaussian envelope with standard deviation $\sigma=6.4$~ns, which is truncated at $\pm 2.5\sigma$. 

Repeating the measurement after scaling the pulse amplitude based on the fit result, we are able to reduce the observed rotation errors [see orange points in Fig.~\ref{fig:fig1}(c)]. The small residual oscillations exhibited by the orange points are not caused by rotation errors captured by our model, and could originate, for example, from off-resonant driving by higher harmonic frequency components, or from uncompensated non-linear distortions of the pulse. The latter effects are beyond the scope of the current work.

To determine the impact of non-linearity for rotation angles smaller than $\pi$, we generalize the technique by using groups of pulses implementing rotation angles that add up to $\pi$. The initial pulse amplitude for these gates is obtained from a proportional reduction with respect to 
the previously calibrated $\pi$-pulse amplitude. 

To calibrate rotation angles up to $\pi/2$, we use a pulse sequence in which each $\pi$-pulse is split into $k$ repetitions of $X_{\pi/k+\varepsilon}$ gates, see the pulse sequence in Fig.~\ref{fig:fig1}(d).  For an $X_{\pi/2}$ rotation, we find an over-rotation error of around $0.4^{\circ}$ [see Fig.~\ref{fig:fig1}(d)], corresponding to an amplitude error of around 0.7 mV. 

Similarly, to measure the non-linearity for rotation angles in the range $\pi/2$ to $\pi$, we apply groups of two $X$ rotations, where the first is a previously calibrated $\pi/k$-gate, and the second is a pulse implementing a rotation of $\pi - \pi/k+\varepsilon$, which is to be calibrated [indicated by the colored box in the pulse sequence above Fig.~\ref{fig:fig1}(e)]. We perform the calibration for an $X_{5\pi/6}$ gate, and find an under-rotation error of around $-1.0^{\circ}$, corresponding to an amplitude error of around $-1.9$ mV. 

For both $X_{\pi/2}$ and $X_{5\pi/6}$, the accumulation of rotation errors is prevented by adjusting the pulse amplitude with the value extracted from the corresponding fit (see orange points).

We reconstruct the non-linear response curve of the qubit drive line by performing the calibration for eleven distinct rotation angles with a gate duration of 15 ns, see Fig.~\ref{fig:fig1}(f, g). This gate duration corresponds to a $\pi$-pulse amplitude of $730 \mathrm{~mV}$ at the output of the AWG, which is around 100 mV below the 1-dB compression point of our upconversion device. Figure~\ref{fig:fig1}(h, i) shows that a simple linear downscaling of the amplitude with respect to the calibrated $\pi$-pulse amplitude introduces a systematic over-rotation error, see the filled purple region in Fig.~\ref{fig:fig1}(i). For rotation angles between 0 and 180 degrees, we model this response with a polynomial correction which preserves the anti-symmetry between positive and negative amplitudes: 
\begin{equation} \label{eq:poly}
    \theta(\Tilde{A}) = 180^{\circ}(1 + b(\Tilde{A}^2-1) + a(\Tilde{A}^4-1))\Tilde{A}, 
\end{equation}
Here, $\Tilde{A}=A/A_{\pi}$ is the pulse amplitude scaled with respect to the amplitude of a $\pi$ rotation, and $a$ and $b$ are fitting parameters capturing the non-linearity of the upconversion circuitry. The degree of the polynomial is chosen heuristically based on the shape of the deviation from the linear scaling shown in Fig.~\ref{fig:fig1}(i). The parameters $a$ and $b$ are determined from a fit to the data in Fig.~\ref{fig:fig1}(h) (orange line), for which we obtain residuals on the order of $0.2^{\circ}$. In Sec.~\ref{sec:cont_xeb}, we use the model in Eq.~(\ref{eq:poly}) to implement arbitrary rotation angles.

In the specific case of our setup, we identify the frequency upconversion device as the dominant source of the non-linear response by comparing the results measured with the $N$-pulse calibration to the ones obtained by rerouting the upconverted drive signal directly to the detection electronics at room temperature (see Appendix~\ref{app:nl_model}). However, not all the observed non-linearity is explained by this measurement, indicating some further source of non-linearity in other parts of the drive line, which could not be avoided by only using this type of calibration. On the other hand, the $N$-pulse calibration method is a calibration tool for the entire microwave drive line that can also be used when the non-linear distortions have a different origin than the one we have identified in our setup.

\section{Impact of non-linearity on short single-qubit gates}
\label{sec:prb}

\begin{figure}
    \centering
    \includegraphics{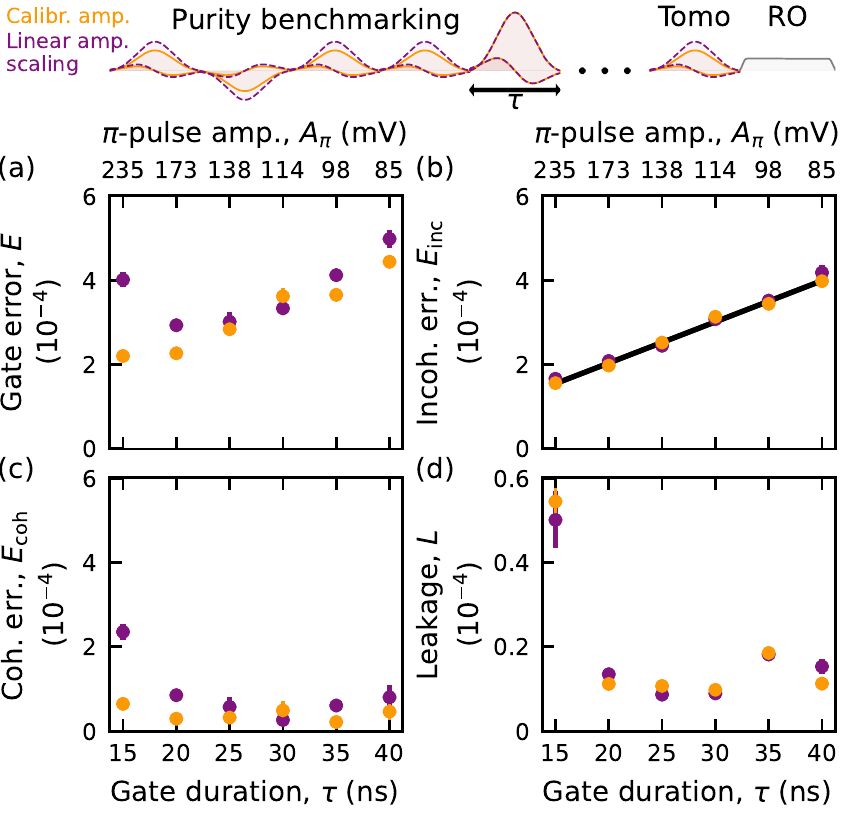}
    \caption{Single-qubit gate performance extracted from Clifford randomized and purity benchmarking (RB/PB). Top: Schematic of a PB sequence with single-qubit gates implemented by linearly scaling the $\pi/2$-pulse amplitude with respect to the amplitude of a $\pi$ rotation (purple), and by calibrating the $\pi/2$-pulse amplitude with the $N$-pulse calibration (orange). In both cases, the $\pi$ pulse is calibrated with the $N$-pulse method. 
    (a)~Total gate error $E$, (b)~incoherent error per gate $E_{\mathrm{inc}}$, (c)~coherent error per gate $E_{\mathrm{coh}}$, and (d)~leakage error per gate $L$ of single-qubit gates as a function of gate duration $\tau$ (bottom axes) and $\pi$-pulse amplitude $A_{\pi}$ (top axes). The black line in (b) is a linear fit to the orange data points. Colors have the same meaning as in the sequence schematic.
    }
    \label{fig:fig2}
\end{figure}

A lower bound on single-qubit gate errors is imposed by the ratio between the gate duration and the coherence time of the system. However, decreasing the gate duration requires an increase of the pulse amplitude. As discussed above, this leads to a larger non-linear distortion [see Fig.~\ref{fig:fig1}(f,g)], and hence, more rotation errors if not calibrated out; but also to possible leakage out of the computational subspace. 

To quantify the impact of rotation errors produced by drive-line non-linearity on the performance of single-qubit gates, we measure the total average gate error with randomized benchmarking (RB)~\cite{Magesan2011,Magesan2012,Epstein2014}. In addition, we distinguish between coherent and incoherent errors using purity bechmarking (PB)~\cite{Feng2016a,Wallman2015a}. Both methods rely on the application of random sequences sampled from the 24 single-qubit Clifford operations, which we decompose into $X$ and virtual $Z$ gates~\cite{McKay2017}. However, for purity benchmarking we apply the tomography pulses $I$ (identity), $X_{\pi/2}$, and $Y_{\pi/2}$ at the end of each sequence to measure the state purity (see the pulse sequence at the top of Fig.~\ref{fig:fig2}). We record 4096 acquisitions of each sequence using single-shot readout, and discriminate between the evolution confined to the qubit computational subspace and leakage into the $f$-state with an average readout assignment error over the three qutrit states of around 4\%. The results are then corrected for readout errors~\cite{Bialczak2010,Krinner2022}. 

We calculate the average $f$-state population for each sequence length and extract the leakage error per gate $L$ [Fig.~\ref{fig:fig2}(d) and Appendix~\ref{app:PRB}]. To estimate the gate errors within the qubit subspace, we then renormalize the populations of the qubit subspace so that they add up to one. The total gate error $E$ [Fig.~\ref{fig:fig2}(a)], and the incoherent error per gate $E_{\mathrm{inc}}$ [Fig.~\ref{fig:fig2}(b)] are extracted independently from the decay of sequence fidelity and state purity, respectively (see Appendix~\ref{app:PRB}). Thus, the difference between $E$ and $E_{\mathrm{inc}}$ provides an estimate of the coherent control errors per gate $E_{\mathrm{coh}}$ in the computational subspace [Fig.~\ref{fig:fig2}(c)]. We stress that, by renormalizing the qubit subspace before estimating the gate errors, $L$ does not contribute to the total gate error $E$.

In Sec.~\ref{sec:np_calib} we found that not correcting for the effects of non-linearity results in around one degree of rotation error for a 32-ns gate, and $3.6^{\circ}$ when the length is decreased to 15 ns. Here we use a different device (device B) with better coherence and smaller $\pi$-pulse amplitude (see Appendix~\ref{app:qbs_used}) to measure the coherent errors introduced when the $\pi/2$-pulse amplitude is scaled linearly relative to the $\pi$-pulse amplitude (purple points in Fig.~\ref{fig:fig2}). By calibrating the $\pi/2$-pulse amplitude with the $N$-pulse method (orange), we show that these errors can be mitigated. In both cases, the amplitude of the $\pi$-pulse is calibrated with the $N$-pulse method. We benchmark the performance of single-qubit gates with lengths ranging from 15 to 40 ns (see the bottom horizontal axis in Fig.~\ref{fig:fig2}), using 50 random purity benchmarking sequences of up to 4096 Cliffords. For each gate length, the pulses are implemented as DRAG pulses with Gaussian envelopes having standard deviations $\sigma$ ranging from 3 to 8 ns which are truncated at $\pm2.5\sigma$. 

For both implementations of the $\pi/2$-pulse, we find an expected linear increase of $E_{\mathrm{inc}}$ with gate duration, see the black line in Fig.~\ref{fig:fig2}(b). At 20~ns, we observe an increase in the coherent errors when using a linear scaling of the amplitude [Fig.~\ref{fig:fig2}(c)]. This increase becomes more pronounced at 15 ns, where the coherent gate errors account for about half of the total gate error [Fig.~\ref{fig:fig2}(a)], while the leakage remains comparatively low at around $6\times 10^{-5}$. With the $N$-pulse calibration we obtain coherence-limited gate errors for all gate lengths, see orange points. This indicates that the measured coherent errors originate primarily from the non-linearity identified in Sec.~\ref{sec:np_calib}, which becomes more pronounced at shorter gate lengths where larger pulse amplitudes are needed to achieve the same rotation (see top axes in Fig.~\ref{fig:fig2}). In our experiment, the coherent errors produced by this effect are up to one order of magnitude larger than leakage into the non-computational states.

\section{IMPACT OF NON-LINEARITY ON ARBITRARY-ANGLE SINGLE-QUBIT GATES} \label{sec:XEB}

\begin{figure}
  \centering
  \includegraphics{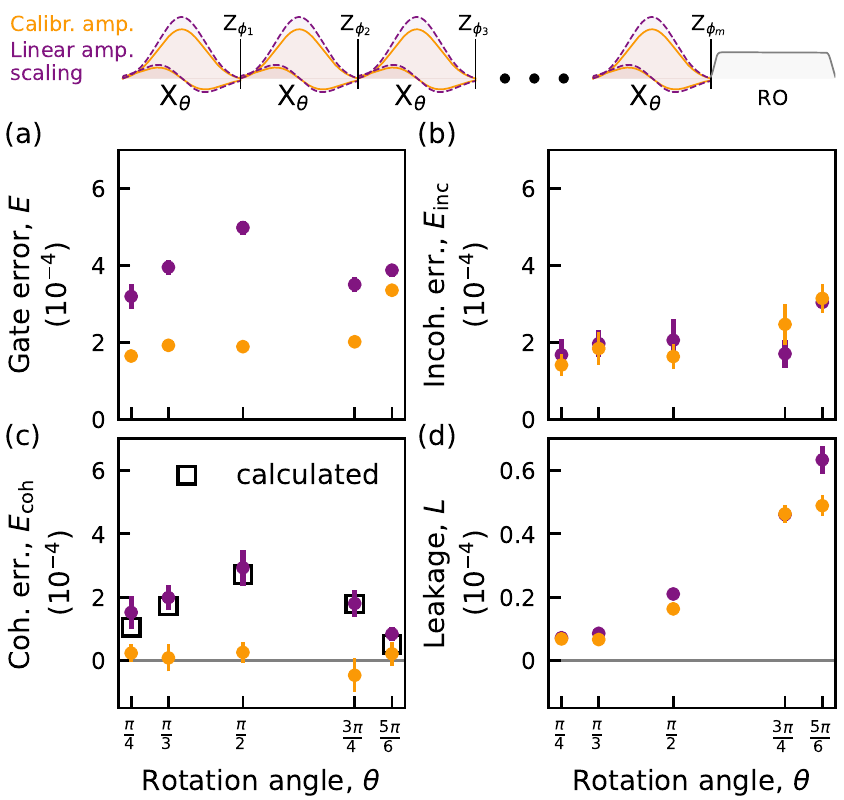}
  \caption{Characterization of arbitrary-angle single-qubit gates using cross-entropy benchmarking (XEB). Top: illustration of an XEB pulse sequence. For each angle $\theta$, the amplitudes of the $X_{\theta}$ gates are obtained either from a linear scaling with respect to the calibrated amplitude of the $X_{\pi}$ gate (purple), or from the $N$-pulse calibration (orange). (a)~Total gate error $E$, (b)~incoherent error per gate $E_{\mathrm{inc}}$, (c)~coherent error per gate $E_{\mathrm{coh}}$, and (d)~leakage error per gate $L$ for single-qubit $X_{\theta}$ gates of various rotation angles $\theta$. The black squares in (c) are calculated from the amplitude corrections obtained with the $N$-pulse calibration. 
  }
  \label{fig:fig3}
\end{figure}

In this section, we show that the $N$-pulse calibration method can be used to implement arbitrary-angle single-qubit gates with coherence-limited performance. Purity benchmarking cannot be used to benchmark arbitrary-rotation single-qubit gates, since these gates are not in the Clifford group. Hence, we use cross-entropy benchmarking (XEB)~\cite{Foxen2020a,Arute2019,Boixo2018a} with the pulse sequence shown at the top of Fig.~\ref{fig:fig3}. Each cycle consists of an $X$ rotation of fixed angle $\theta$ followed by a virtual $Z_{\phi_i}$ gate, with $\phi_i$ sampled fully randomly from a uniform distribution between 0 and $2\pi$. We generate $K=200$ random sequences of up to 4096 cycles and measure in the $Z$ basis using single-shot readout with 4096 acquisitions.

In cross-entropy benchmarking, the cross-entropy is used as a metric to compare the output qubit-state distribution after $K$ random sequences to the one calculated assuming ideal gates~\cite{Foxen2020a}. By then fitting the exponential decay of this cross-entropy fidelity versus the number of cycles in the sequence, we extract the average error $E$ of the $X_{\theta}$ gate (see Appendix~\ref{app:XEB}). In addition, the ratio of the variance of the measured distribution and that of the Porter-Thomas distribution allows us to extract the average purity of the final qubit state over the distribution of random sequences~\cite{Arute2019}. This quantity also decays exponentially with the number of cycles, from which we extract the average incoherent errors per gate, $E_{\mathrm{inc}}$. Then we calculate the coherent errors per gate $E_{\mathrm{coh}}$ as the difference between $E$ and $E_{\mathrm{inc}}$. Finally, as also done for purity benchmarking, we use the measured $f$-state population to estimate the average leakage error per gate $L$, and renormalize the qubit subspace before estimating the gate errors. Thus, $L$ does not contribute to the total gate error $E$.

To minimize the incoherent errors, we choose the shortest, 15-ns gate in Fig.~\ref{fig:fig2}, which is also maximally sensitive to errors caused by the non-linearity of the drive-line components. We benchmark the performance of $X$ rotations with angles $\pi/4$, $\pi/3$, $\pi/2$, $3\pi/4$, $5\pi/6$ using the $N$-pulse calibration method, and obtain coherence-limited gate errors in the computational subspace of around $2.0(1)\times10^{-4}$ for all rotation angles [see orange points in Fig.~\ref{fig:fig3}(a)]. Thus, we observe no coherent errors within error bars.

For reference, we repeat the measurements for the case where the amplitude is scaled linearly with respect to the calibrated $\pi$-pulse amplitude (see purple points), and we find a dependence of the control errors on the rotation angle [Fig.~\ref{fig:fig3}(c)], which is similar to the curve in Fig.~\ref{fig:fig1}(i). From the $N$-pulse calibration of each rotation angle performed before the corresponding XEB measurement, we calculate the expected control errors per gate, and find good agreement with the data in Fig.~\ref{fig:fig3}(c), see hollow black squares.

With both implementations, the incoherent errors per gate remain constant across all measured angles [Fig.~\ref{fig:fig3}(b)], as expected for a fixed gate duration. However, the leakage shows a small increase with rotation angle from around $2\times10^{-5}$ for $\pi/4$ to around $6\times10^{-5}$ at $5\pi/6$ [Fig.~\ref{fig:fig3}(d)], which is expected as the drive amplitude becomes larger. Nevertheless, the leakage is smaller by one order of magnitude than the coherent errors introduced by the non-linearity, signalling the latter as the main limitation on overall gate performance.

\section{CONTINUOUSLY PARAMETERIZED SINGLE-QUBIT GATES} \label{sec:cont_xeb}

\begin{figure}
  \centering
  \includegraphics{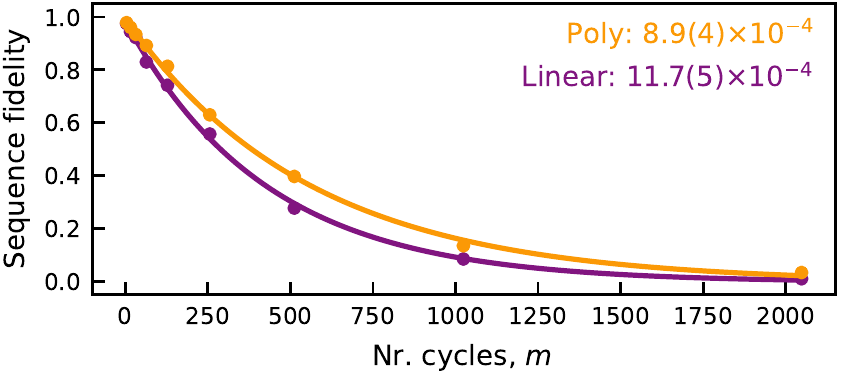}
  \caption{Cross-entropy benchmarking (XEB) of random-angle single-qubit $X$ rotations. The data points show the sequence fidelity, and the line is an exponential fit to the data.  The legend indicates the total gate error obtained when using linear (purple) and polynomial (orange) scaling for calculating the amplitudes of the $X$ pulses.}
  \label{fig:fig4}
\end{figure}

Finally, we show that an individual calibration of every rotation used in an algorithm is not required to achieve coherence-limited gates. To this end, we implement a continuously parameterized gate set using a polynomial scaling of the pulse amplitude obtained from the fit of the data in Fig.~\ref{fig:fig1}(h) to the model in Eq.~(\ref{eq:poly}).

To benchmark continuously parameterized single-qubit gates, we replace the fixed-angle $X_{\theta}$ gates in the protocol described Sec.~\ref{sec:XEB} by $X_{\theta_i}$ gates, where in each cycle, the angles $\theta_i$ are sampled fully randomly from a uniform distribution between 0 and $\pi$. We use 400 random sequences of up to 2048 cycles, and measure each sequence in the $Z$ basis using single-shot readout with 8192 acquisitions. We discriminate between the computational subspace and leakage, and process the data as described in Sec.~\ref{sec:XEB}.

When the pulse amplitudes are scaled with the polynomial function, we measure coherence-limited random-angle $X$ gates with an average error of $8.9(4)\times 10^{-4}$ on device A, see the orange data points in Fig.~\ref{fig:fig4} and the exponential fit (line). With a linear amplitude scaling (purple points and line), we find around $3.8\times 10^{-4}$ of coherent control errors. This number matches the mean value of the gate errors calculated from the rotation errors in Fig.~\ref{fig:fig1}(h, i).

\section{CONCLUSION}

In this work, we have presented the \emph{$N$-pulse calibration} method for pulse amplitudes, which amplifies small rotation errors by the repeated application of gates. Such errors can originate from signal distortions introduced by non-linear components that are essential for processing microwave signals, such as the frequency upconversion circuitry used in our experiment. By correcting the rotation errors, we demonstrated -- using randomized and purity benchmarking -- that this non-linearity can be the dominant source of single-qubit coherent gate errors, and that the errors produced by the non-linearity can be up to one order of magnitude larger than leakage into the $f$-state. In particular, in our setup using superconducting transmon qubits, we found up to several degrees of over-rotation error for the shortest gates we have investigated (15 ns), which accounts for half the total gate error. Finally, by scaling the pulse amplitude based on a simple polynomial model of the non-linear response of our control line, we showed that we can implement a continuous family of single-qubit gates with coherence-limited performance for any rotation angle, as characterized with cross-entropy benchmarking.

Our method demonstrates that a sample-by-sample correction of the pulse shape is not needed to mitigate the control errors introduced by the non-linearity and obtain coherence-limited gates. Moreover, the $N$-pulse technique is an \emph{in-situ} calibration measurement which can be used to compensate for any source of non-linearity. However, while our protocol tackles rotation errors, it is not highly sensitive to other types of control errors, such as leakage, crosstalk, or phase errors which may arise from off-resonant driving by spurious frequency components. Nevertheless, similar error amplification techniques, such as the ones used in Ref.~\cite{Chen2016,Lucero2010}, can be tailored to address these error mechanisms as well. 

This work demonstrates that as qubit coherence times improve in the near future, the characterization and mitigation of coherent single-qubit control errors is crucial for further improving the fidelity of these gates. The $N$-pulse calibration is an essential step in this direction. Our method is applicable to any quantum computing platform suffering from non-linear distortions of the drive signals. In particular, the $N$-pulse calibration method is an essential tool for architectures that are compatible with large driving amplitudes, such as superconducting circuits with larger anharmonicity~\cite{Ficheux2021,Somoroff2021,Liu2021o,Hyyppa2022a}.

\section*{Acknowledgments}

The authors acknowledge the contributions of Luca Hofele to device calibration and characterization.

The authors acknowledge financial support by the Office of the Director of National Intelligence (ODNI), Intelligence Advanced Research Projects Activity (IARPA), via the U.S. Army Research Office grant W911NF-16-1-0071, by the EU Flagship on Quantum Technology H2020-FETFLAG-2018-03 project 820363 OpenSuperQ, by the National Centre of Competence in Research Quantum Science and Technology (NCCR QSIT), a research instrument of the Swiss National Science Foundation (SNSF), by the SNFS R'equip grant 206021-170731, by the Baugarten Foundation and the ETH Zurich Foundation, by the EU program H2020-FETOPEN project 828826 Quromorphic, and by ETH Zurich. 
The views and conclusions contained herein are those of the authors and should not be interpreted as necessarily representing the official policies or endorsements, either expressed or implied, of the ODNI, IARPA, or the U.S. Government.

\section*{Author Contributions}

S.L. and Q.F. planned the experiments, and S.L. performed the main experiment and analyzed the data.
A.R. and F.S. designed the devices, and D.C., A.R., G.J.N., M.B.P., A.F., M. K., and J.C.B. fabricated the devices.
S.L., C.H., N.L., and A.R., developed the experimental software framework, and S.L. developed the control and calibration software routines used in this work.
S.L., A.R., C.H., and J.H. designed and built elements of the room-temperature setup, and S.L., Q.F., N.L., C.H., and J.H. maintained the experimental setup.
S.L., Q.F., and N.L. characterized and calibrated the devices and the experimental setup.
S.L. prepared the figures for the manuscript, and S.L. and Q.F. wrote the manuscript with inputs from the co-authors.
C.E. and A.W. supervised the work.

\newpage
\begin{appendix}

\section{DEVICE PARAMETERS} \label{app:qbs_used}
 
In Table~\ref{tab:qubit_measured_parameters} below, we summarize the properties of the qubits on the two devices (A and B) with which the experiments were performed, and indicate for which figure of the main text they were used.

\begin{table}[H]
    \centering
    \begin{tabular}{lrr}
    \toprule
              &       Device A    &       Device B \\
    \midrule
     
    Qubit idle frequency, $\omega_Q/2\pi$ (GHz) & 6.257 & 4.640 \\ 
    Qubit anharmonicity, $\alpha/2\pi$ (MHz) & -153 & -183 \\
    Lifetime, $T_1$ ($\mu$s)    & 12.3 &  60.9 \\
    Ramsey decay time, $T_2^*$ ($\mu$s)  & 9.86 &  55.3  \\
    Echo decay time, $T_2^\mathrm{e}$ ($\mu$s)  & 14.7 & 67.3  \\
    15-ns $\pi$-pulse amplitude, $A_{\pi}$ (mV)  & 730 & 235 \\
    Three-state readout error, $\epsilon_{\mathrm{RO}}^{(3)}$ (\%)  & 3 & 4 \\    
    \bottomrule
    
    \end{tabular}
    \label{tab:qubit_measured_parameters}
    \caption{Qubit parameters, and their coherence, drive, and readout properties. The measurements in Fig.~\ref{fig:fig1} and Fig.~\ref{fig:fig4} were performed on device A, while those in Fig.~\ref{fig:fig2} and Fig.~\ref{fig:fig3} were done on device B.}    
\end{table}

\section{AMPLITUDE CORRECTION WITH ERROR AMPLIFICATION} \label{app:fit}

The excited state population of the qubit after the application of a series of $N$ $\pi$-pulses [see Fig.~\ref{fig:fig1}(c-e)] is fitted to a master equation model to extract the error in the rotation angle. 

We model our measurement starting from the time-independent Hamiltonian $H=\Omega \sigma_x/2$ describing rotations around the $x$ axis of the Bloch sphere at the Rabi rate $\Omega$. The time evolution of the $y$ and $z$ components of the Bloch vector are captured by the Lindblad equation
\begin{align} \label{eq:me}
    \frac{d\rho(t)}{dt} =& -i [ H,\rho(t)] + \frac{\Gamma_{\phi}}{2} (\sigma_z \rho(t)\sigma_z - \rho(t))~+ \notag \\  
    & \Gamma_1 \bigg( \sigma_{-}\rho(t)\sigma_+ - \frac{\sigma_+\sigma_{-}\rho(t) + \rho(t)\sigma_+\sigma_{-}}{2} \bigg),
\end{align}
with $\sigma_+=|1\rangle \langle 0|$, $\sigma_-=|0\rangle \langle 1|$, the energy relaxation rate of the qubit $\Gamma_{1}=1/T_1$, and the dephasing rate $\Gamma_{\phi}=1/T_2-1/(2T_1)$.

From Eq.~(\ref{eq:me}), we obtain the time evolution of the Bloch vector components of the qubit state in the $yz$ plane as~\cite{Ficheux2018a}
\begin{widetext} 
    \begin{equation} \label{eq:time_evo}
        \begin{pmatrix}
        \Tilde{y}(t) \\
        \Tilde{z}(t) 
        \end{pmatrix}    
        = e^{-(3\Gamma_1+2\Gamma_{\phi})\frac{t}{4}}
        \begin{pmatrix}
        
        \mathrm{cos}(\nu t) + \mathrm{sin}(\nu t) \frac{\Gamma_1-2\Gamma_{\phi}}{4\nu} & \frac{\mathrm{sin}(\nu t)}{\nu}\Omega \\
        -\frac{\mathrm{sin}(\nu t)}{\nu}\Omega & \mathrm{co}s(\nu t) - \mathrm{sin}(\nu t) \frac{\Gamma_1-2\Gamma_{\phi}}{4\nu} 
        
        \end{pmatrix}
        \begin{pmatrix}
        \Tilde{y}(0) \\
        \Tilde{z}(0) 
        \end{pmatrix},      
    \end{equation}
\end{widetext}
where
\begin{align}
    \Tilde{y}(t) &= y(t) + \frac{2\Omega \Gamma_1}{\Gamma_1(\Gamma_1+2\Gamma_{\phi})+2\Omega^2}, \\
    \Tilde{z}(t) &= z(t) + \frac{\Gamma_1(\Gamma_1+2\Gamma_{\phi})}{\Gamma_1(\Gamma_1+2\Gamma_{\phi})+2\Omega^2}, \\ 
    \nu &= \sqrt{\Omega^2 - \frac{(\Gamma_1 - 2\Gamma_{\phi})^2}{16}},
\end{align}
and $\Tilde{y}(0), \Tilde{z}(0)$ are the coordinates at the start of the evolution. 

We apply Eq.~(\ref{eq:time_evo}) to the qubit ground state ($y(0)=0$, $z(0)=-1$), and then continue to apply this equation to the qubit state after 
each pulse in the $N$-pulse calibration measurement. We estimate the final qubit excited state population as $p_e=(1+z(t))/2$. Denoting the gate duration by $\tau$, we express $\Omega=\alpha \pi/\tau$, and extract the dimensionless fraction of a $\pi$ rotation $\alpha$ from a fit of the measured data to $p_e$. Then we calculate the pulse amplitude as $A=\alpha A_{\pi}$, where $A_{\pi}$ is the amplitude of a $\pi$ rotation.

\section{SIGNAL NON-LINEARITY DUE TO ELECTRONIC COMPONENTS} \label{app:nl_model}

\begin{figure*}[t]
    \centering
    \includegraphics{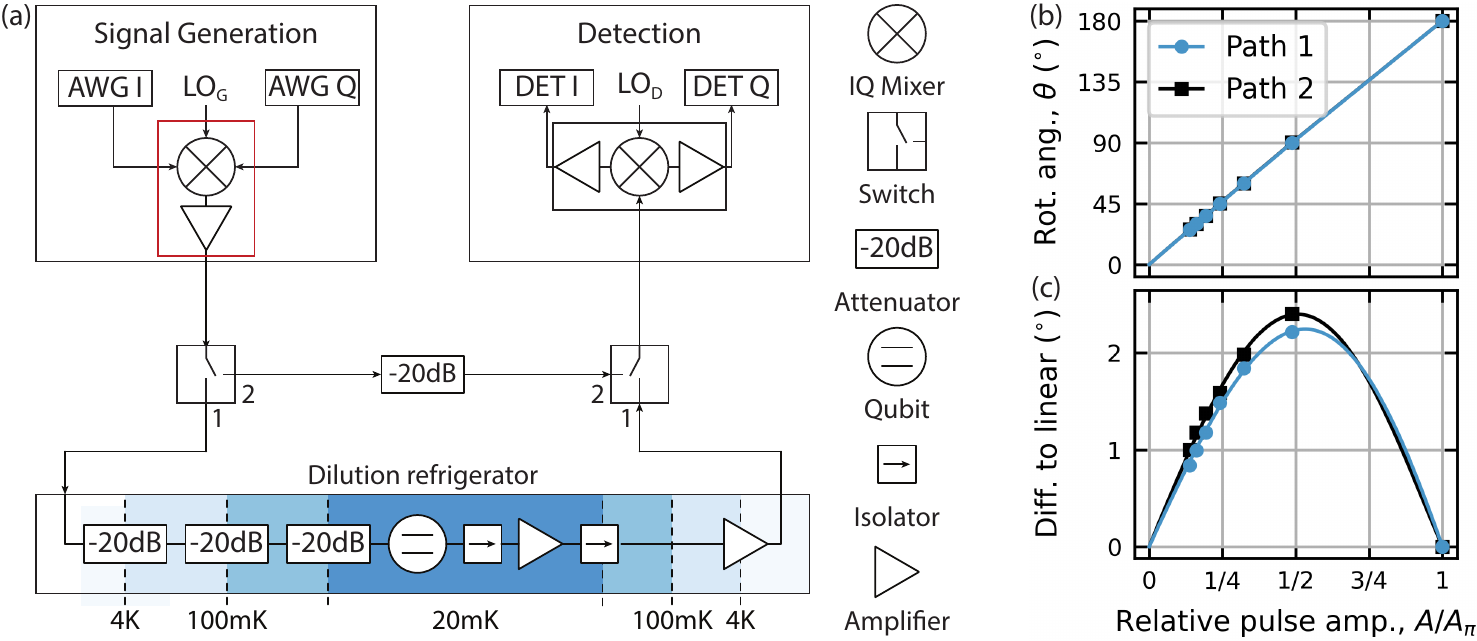}
    \caption{(a) Simplified schematic of the room temperature and measurement setup used to operate the qubits. Intermediate frequency pulses are generated with an AWG and upconverted to the GHz range using the conversion device indicated in red. This device contains an IQ mixer and an amplifier, and is the main source of non-linearity in our drive line. We configure a switch to either send the pulses into the dilution refrigerator via the qubit drive line (path 1), or directly into the detection chain consisting of another IQ mixer, which downconverts the pulses into the frequency band of the detection device (path 2). (b) Measured qubit rotation angle as a function of applied pulse amplitude relative to the $\pi$-pulse amplitude. The blue points were measured with the $N$-pulse calibration via path 1, and the black points are obtained from the amplitudes of the demodulated pulses measured via path 2 and converted to the same units. (c) Difference between the rotation angles in (b) and a linear interpolation between the last point and the origin. The lines in (b) and (c) are fits to the fifth-order polynomial in Eq.~(\ref{eq:poly}) of the main text. 
    }
    \label{fig:app_fig1}
\end{figure*}

The response curves shown schematically in Fig.~\ref{fig:fig1}(b) and measured in Fig.~\ref{fig:fig1}(f-i) are typical for electronic components such as mixers or amplifiers, which become non-linear above a certain input power~\cite{Pedro2018}. 

Here we verify that the non-linearity observed with the $N$-pulse calibration originates mostly from the control electronics used for driving the qubits. We perform the measurements on a different setup than the ones presented in the main text, but with nominally identical control electronics. We use slightly longer 50-ns DRAG pulses, with a $\sigma=10$ ns standard deviation of the Gaussian envelope truncated at $\pm 2.5\sigma$. The pulses are generated at an intermediate frequency of -100 MHz before upconversion to the qubit frequency using a circuit which consists of an IQ mixer and amplifiers [see the red outline in the schematic in Fig.~\ref{fig:app_fig1}(a)]. We choose to either apply the pulses to the qubit and perform the $N$-pulse calibration [path 1 in Fig.~\ref{fig:app_fig1}(a)], or we directly downconvert the pulses to an intermediate frequency of 200 MHz before detecting the quadratures of the signal (path 2). The signal for path 2 is attenuated by -20 dB to ensure that the downconversion mixer and subsequent amplifiers are operated in their linear regime. We then demodulate the signal to DC, filter out spurious frequency components larger than the spectral width of the pulse, and extract the amplitudes of the resulting pulses. This data is then converted to a rotation angle [see black squares in Fig.~\ref{fig:app_fig1}(b)] for better comparison to the results obtained with the $N$-pulse calibration via path 1 (blue circles).

We find that the non-linear scaling of the Rabi angle with the pulse amplitude identified with the $N$-pulse calibration technique originates mostly from our frequency conversion device, see Fig.~\ref{fig:app_fig1}(c), where we plot the difference between the measured data points in Fig.~\ref{fig:app_fig1}(b) and a linear interpolation between the origin and the last data point. However, the results obtained via path 2 systematically overestimate the ones measured using the qubit by about $0.2^{\circ}$ at an amplitude ratio of 0.5, showing that the room-temperature calibration presented here cannot fully replace the \emph{in-situ} calibration presented in the main text.

\section{PURITY BENCHMARKING} \label{app:PRB}

\begin{figure}
    \centering
    \includegraphics{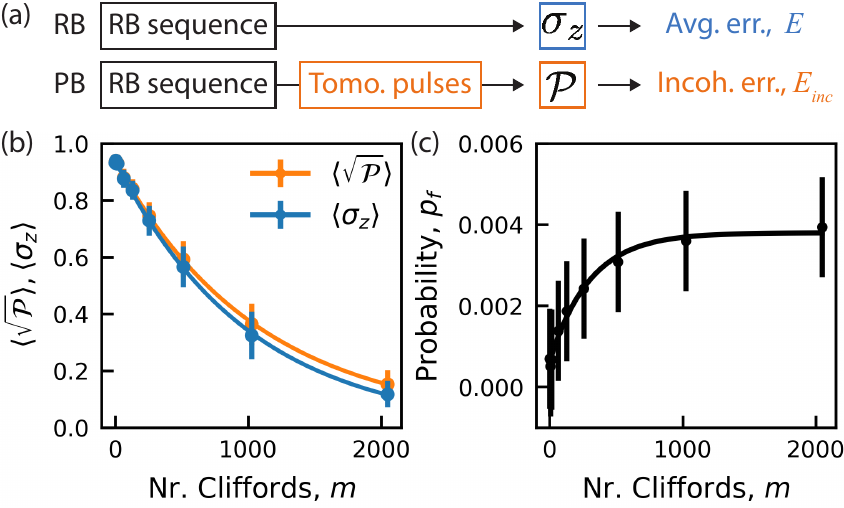}
    \caption{RB and PB with the 40-ns single-qubit gate shown in Fig.~\ref{fig:fig2}. (a) Average expectation value of the $\sigma _z$ operator (blue), and the average of the square root of the sequence purity $\sqrt{\mathcal{P}}$ (orange) as a function of the number of Clifford gates. The lines are fits to the exponential decay functions in Eq.~(\ref{eq:rb_fits}). (c) $f$-state population as a function of the number of Clifford gates (points) with a fit to the rate equation Eq.~(\ref{eq:leak_rate}) (line). The error bars in (b) and (c) represent one standard deviation of the distribution of outcomes of the random sequences for a fixed number of Clifford gates.}
    \label{fig:app_fig2}
\end{figure}

In order to distinguish between incoherent and control errors, we use a simple extension of the randomized benchmarking (RB) protocol known as purity benchmarking (PB). Purity benchmarking consists of performing tomography of the qubit state vector after each random sequence of Clifford gates~\cite{Feng2016a,Wallman2015a}. This is illustrated by the schematic in Fig.~\ref{fig:app_fig2}(a), where the orange block consists of the single-qubit tomography pulses $I$ (identity), $X_{\pi/2}$, and $Y_{\pi/2}$. We estimate the purity of the state after each random sequence as
\begin{equation}
    \mathcal{P} = \langle \sigma_x \rangle ^2 + \langle \sigma_y \rangle ^2 + \langle \sigma_z \rangle ^2,
\end{equation}
and compute the mean of this quantity over the random sequences for a fixed number of Cliffords.

In Fig.~\ref{fig:app_fig2}(b) we show the exponential decay of the $\langle \sigma_z \rangle$ operator (blue) from the RB measurement of the 40-ns gate in Fig.~\ref{fig:fig2} of the main text calibrated with the $N$-pulse technique. We compare this decay to the decay of $\langle \sqrt{\mathcal{P}} \rangle$ (orange) from the corresponding PB experiment. In the absence of control errors, the two curves should coincide. 

For RB and PB, respectively, we fit to
\begin{align} \label{eq:rb_fits}
    \langle \sigma_z \rangle (m) &= A\alpha ^m + B,  \\
    \langle \mathcal{P} \rangle (m) &= A'u^m + B', \notag
\end{align}
where $m$ is the number of Cliffords, $\alpha$ is the depolarization parameter~\cite{Magesan2011}, $u$ is the unitarity~\cite{Wallman2015a}, and $A$, $A'$, $B$, $B'$ are fit parameters capturing state preparation and measurement (SPAM) errors. The total average error per gate and the average incoherent error per gate, respectively, are then calculated as,
\begin{align}
    E &= \frac{1-\alpha}{2N},  \\
    E_{\mathrm{inc}} &= \frac{1-\sqrt{u}}{2N}, \notag
\end{align}
where $N=1.125$ is the average number of physical gates per Clifford in the $HZ$ decomposition~\cite{McKay2017}. The virtual $Z$ gates are implemented in software by a qubit phase update, and hence it has perfect fidelity. Subtracting these two quantities gives an estimate of the coherent control error per gate, $E_{\mathrm{coh}}=E-E_{\mathrm{inc}}$. 

The analysis described above is performed after discriminating between the qubit subspace and leakage, and normalizing the qubit populations so that they add up to one. In Fig.~\ref{fig:app_fig2}(c), we show the average $f$-state population as a function of the RB sequence length (points). We fit this data to the following rate equation~\cite{Chen2016}
\begin{equation} \label{eq:leak_rate}
    p_f(m) = \frac{l}{l+s}\Big(1 - e^{-(l+s)m}\Big) + p_0e^{-(l+s)m},
\end{equation}
from which we find the average leakage per gate, $L=l/N$. Here, $s$ is the decay rate into the computational subspace, and $p_0$ is the initial $f$-state population.

From the data shown in Fig.~\ref{fig:app_fig2}, we extract a leakage rate per gate of around $1.1(1)\times 10^{-5}$, and a total average gate error of $4.4(2)\times 10^{-4}$, which is limited by decoherence at around $4.1(1)\times 10^{-4}$.

\section{CROSS-ENTROPY BENCHMARKING}\label{app:XEB}

\begin{figure}
    \centering
    \includegraphics{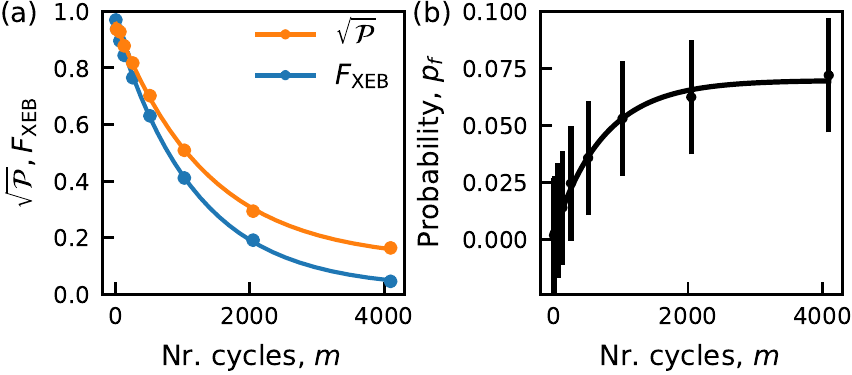}
    \caption{XEB with the 15-ns single-qubit $X_{5\pi/6}$ gate shown in Fig.~\ref{fig:app_fig2}. (a) Average XEB sequence fidelity $F_{\mathrm{XEB}}$ (blue) and the square root of the average sequence purity $\sqrt{\mathcal{P}}$ (orange) as a function of the number of cycles. The lines are fits to the exponential decay functions in Eq.~(\ref{eq:xeb_fits}). (b) $f$-state population as a function of the number of cycles (points) with a fit to the rate equation Eq.~(\ref{eq:leak_rate}) (line). The error bars represent one standard deviation of the distribution of outcomes of the random sequences for a fixed number of cycles.}
    \label{fig:app_fig3}
\end{figure}

To benchmark the performance of gates that do not belong to the Clifford group, such as arbitrary-rotation single-qubit gates, we use the cross-entropy benchmarking technique~\cite{Foxen2020a,Arute2019,Boixo2018a}. In the XEB protocol, sequences of random cycles of gates are applied to a system prepared in a chosen state (typically the ground state). The only requirement to take into consideration when choosing the cycle design is that sequences with enough cycles should fully explore the Hilbert space of the system~\cite{Boixo2018a}. 

The average sequence fidelity is estimated by comparing the distribution of measured outcomes $Q_{\mathrm{meas}}$ to the calculated, ideal distribution $Q_{\mathrm{ideal}}$ using the cross-entropy as a metric of overlap. The cross-entropy between two distributions $Q$ and $R$ is given by
\begin{equation}
    S(Q,R) = -\sum_i q_i \mathrm{ln}(r_i),
\end{equation}
where $q_i,r_i$ are samples from the distributions $Q,R$. For a fixed number of cycles, we estimate the average XEB sequence fidelity of a distribution of random sequence outcomes as~\cite{Foxen2020a}
\begin{equation}
    F_{\mathrm{XEB}} = \frac{ S(Q_{\mathrm{incoh}}, Q_{\mathrm{ideal}}) -  S(Q_{\mathrm{meas}}, Q_{\mathrm{ideal}}) }{ S(Q_{\mathrm{incoh}}, Q_{\mathrm{ideal}}) - S(Q_{\mathrm{ideal}}, Q_{\mathrm{ideal}}) }.
\end{equation}
Here, $Q_{\mathrm{incoh}}$ is the incoherent distribution where all the bit strings have an equal probability of $1/2^n$ (with $n=1$ for one qubit), and $Q_{\mathrm{meas}}$ and $Q_{\mathrm{ideal}}$ are the measured and calculated distributions of probabilities $(p_g,p_e)_i$ describing the qubit state at the end of each random sequence $i$. To constrain $F_{\mathrm{XEB}}$ between 0 and 1, we normalize $Q_{\mathrm{incoh}}$, $Q_{\mathrm{meas}}$, and $Q_{\mathrm{ideal}}$ by the total number of random sequences for a fixed number of cycles. 
  
From the variance of the measured probability distribution, we also estimate the average purity of the final qubit state as~\cite{Arute2019}
\begin{equation}
    \mathcal{P} = \frac{\mathrm{Var}(Q_{\mathrm{meas}})}{\mathrm{Var}(Q_{\mathrm{PT}})},
\end{equation} \vspace{0.01cm}

\noindent where $\mathrm{Var}(Q_{\mathrm{PT}})=(2^n-1)/(2^{2n}(2^n+1))$ is the variance of the Porter-Thomas distribution, with $n=1$ for a single qubit.

The points in Fig.~\ref{fig:app_fig3}(a) show the exponential decays of the XEB fidelity (blue) and the square root of the sequence purity (orange) for the purple point in Fig.~\ref{fig:fig3} of the main text, corresponding to an $X$ gate of angle $5\pi/6$. The lines are fits to
\begin{align} \label{eq:xeb_fits}
    F_{\mathrm{XEB}}(m) &= A\bigg(1-\frac{2^{2n}}{2^{2n}-1}\frac{2^n}{2^n-1}E\bigg)^m + B, \notag \\
    \sqrt{\mathcal{P}}(m) &= A'\bigg(1-\frac{2^{2n}}{2^{2n}-1}\frac{2^n}{2^n-1}E_{\mathrm{inc}}\bigg)^m + B',
\end{align}
from which we estimate the total average errors per cycle $E$, and the average incoherent errors per cycle $E_{\mathrm{inc}}$. Here $m$ is the number of cycles, and $A, A', B, B'$ are fit parameters capturing the state preparation and measurement (SPAM) errors. We then calculate the amount of coherent control errors per cycle as $E_{\mathrm{coh}}=E-E_{\mathrm{inc}}$. Since both cycle designs chosen for this work (see Fig.~\ref{fig:fig3} and Fig.~\ref{fig:fig4}) contain only one physical gate, $E$, $E_{\mathrm{inc}}$, and $E_{\mathrm{coh}}$ quantify the average errors per gate. From the data set in  Fig.~\ref{fig:app_fig3}(a), we find a total average gate error of $6.3(3)\times 10^{-4}$, with a contribution from decoherence errors of $5.6(2)\times 10^{-4}$, giving around $8(3)\times 10^{-5}$ control errors.  

The analysis described above is performed after discriminating between the qubit subspace and leakage, and normalizing the qubit populations so that they add up to one. In Fig.~\ref{fig:app_fig3}(b), we show the average $f$-state population (points) as a function of the XEB sequence length, and a fit to the rate equation in Eq.~(\ref{eq:leak_rate}) (line), from which we extract an average leakage rate per gate of around $9.5(7)\times 10^{-5}$.

\end{appendix}

\end{document}